# A Generic Framework For Capturing Reliability in Cyber-Physical Systems

Nazakat Ali, Manzoor Hussain, YoungJae Kim and Jang-Eui Hong[*]
Department of Computer Science, Chungbuk National University, Republic of South Korea
* jehong@chungbuk.ac.kr

## ABSTRACT

Cyber-Physical systems solve complex problems through their tight integration between the physical and computational components. Therefore, the reliability of a complex system is the most critical requirement for the cyber-physical system because an unreliable system often leads to service disruption, property damage, financial loses and sometimes lead to fatality. In order to develop more reliable CPS, this paper proposes a generic framework for reliability modeling and analysis for our ongoing work on cyber-physical systems. The proposed architecture is comprising of physical elements such as sensors and actuators, communication network, computation and control units, and a reliability model. The reliability model covers almost all possible reliabilities required to a general cyber-physical system. Furthermore, we demonstrate the proposed framework with an illustrative example by using different reliability values from "offshore and onshore reliability data" library.

• **Framework→ Generic framework**

## 1. INTRODUCTION

Cyber-Physical Systems (CPSs) have emerged as a new engineering technique which is used to solve complex problems in various fields of information technology such as smart buildings, smart grids, self-driving vehicles, smart manufacturing systems, smart transportation, oil and gas industries, nuclear power plants and avionics [1]. CPSs are composed of physical components such as sensors and actuators, communication network, and cyber components such as computation and control, and monitoring unit. The cyber and physical components are highly intertwined to achieve a specific task. Reliability of CPSs has become a critical concern for researchers because a single failure or a malfunction in the system may lead to property damage, financial loss or even loss of human lives.  Therefore, CPSs are not designed to work in a controlled environment, they must be designed to operate in a robust manner to handle unpredicted conditions and adaptable to system failures [1].

In CPSs, sensors are utilized to collect environmental or system data and transmit this data through a network to central computation and control unit for further processing [2]. The information can be extracted from the processed data and then change in the environment can be responded through actuators.

It is hard to make CPSs reliable. The modern CPSs have become more and more complex due to incorporating components from different providers with different interfaces, physical operation characteristics, and real-time sensing especially for mission-critical online system e.g. smart grids and building management systems need continuous uptime [3]. In CPSs, erroneous data is a major problem when it comes to system reliability which process a large amount of real-time data on the fly.

In this paper, we propose an architecture for a general CPS which includes an environmental (physical) layer, communication and network layer, and computation and control layer (cyber). The environmental layer has sensor module and actuator module. The cyber layer includes computation and control module and reliability model. The computation and control module is composed of hardware and software components. The software and hardware have a strong interaction with each other in order to complete a task. The reliability model is proposed to measure the CPS reliability. The reliability model has sensor reliability, actuator reliability, communication reliability, data reliability, and computation and control unit reliability which includes hardware reliability, software reliability, and software-hardware interaction reliability.

In summary, we make following contributions:
- First, we propose a generic framework for capturing reliability in CPS includes data reliability.
- Second, we demonstrate the proposed framework with an illustrative example by using different reliability values from a "offshore and onshore reliability data" library [25].
- Software-hardware interaction is critical in computing CPS reliability. Therefore, we also present a survey on how software fails due to hardware failure (software-hardware interaction), and how hardware (hardware-software interaction) fails due to interaction with a faulty software.

The rest of paper is organized as follow: Section 2 discusses the reliability modeling for CPS. Section 3 presents an illustrative example to explain our proposed architecture. In Section 4, we investigate related work and Section 5 concludes this study.

## 2. MODELING CPS WITH VARIABILITY

We propose CPS architecture in order to model its reliability characteristics. CPS is consisting of cyber space and physical (environment) space. Physical space and cyber space communicate through a communication network. In our CPS architecture, the CPS is proposed to have reliability model as well because proper functioning of complex systems like CPS is vital due to fact that a single failure in CPS may lead to catastrophic consequences such as financial loss, and probably human death.

The proposed architecture as shown in Fig. 1 has three layers: 1) physical layer, which includes the sensor module and actuator module, 2) communication network layer, where the communication medium between physical and cyber space is defined. The medium can be a wireless network, a wired network or a hybrid communication network, 3) computation and control or cyber layer which includes computation and control module and a reliability model. The computation and control module is consists of software and hardware components. The hardware and software have a strong interaction with each other in order to complete a task. By analyzing modules of the target CPS, we have defined a reliability model which should be included in CPS.

The sensor module of our CPS architecture may consist of a group of sensors that monitor the physical (environment) world and capture the data and transmit it to the cyber world. The actuator module of our CPS architecture may consist of a group of actuators to reflect the response for change in the environment. The data captured through sensors is transmitted to cyber world (computation and communication module) through communication network. The network performs two-way communication between cyber and physical world such as sensors and actuators using Machine-to-Machine (M2M) communication. The M2M is considered to be highly reliable and low latency link for data transmission [30]. The computation and control module of cyber layer is composed of software and hardware components. The software component may include data processing component to process the information received from sensors and sends that status information to the control algorithms component of software to generate the required control signal for actuators. The monitoring component maintains historical data of all the components in order to monitor the health of associated components. The actuators respond to the environment upon receiving the required control signals.

In order to measure the reliability of a CPS, we propose a reliability model for the CPS. The reliability model is consisting of sensors' reliability, to ensure the behavior of sensors, actuators' reliability, to ensure the behaviour of actuators, communication reliability, to ensure the network is working fine, data reliability to ensure the consistency and completeness of input data from sensors. The computation and control reliability is consists of hardware reliability, software reliability, and the software-hardware interaction reliability.

### 2.1 *Reliability Modeling and Analysis for CPS*

For safety-critical systems, continuous availability is an essential requirement and software reliability is one of important components of continuous system availability. Relia-bility is defined as the probability that software or hardware will perform its intended task for a specific time under certain conditions [6]. Reliability is related with time to failure, or its reciprocal, and failure rate. The reliability of whole CPS is determined by computing the reliability of each component. In CPS, its components are designed to work independently. However, in order to achieve a common goal, the components have to work collectively and correctly. Consequently, the failure of a single component may lead to failure of whole system. Therefore, the actual CPS has a series of configurations in terms of its reliability. The reliability of a system component is computed by the probability of successful functioning $P(X_i)$ of that component independently. The reliability of entire system $R_{sys}$ can be the intersection of the probability $P(X_i)$ of each system component, represented as equation (1).

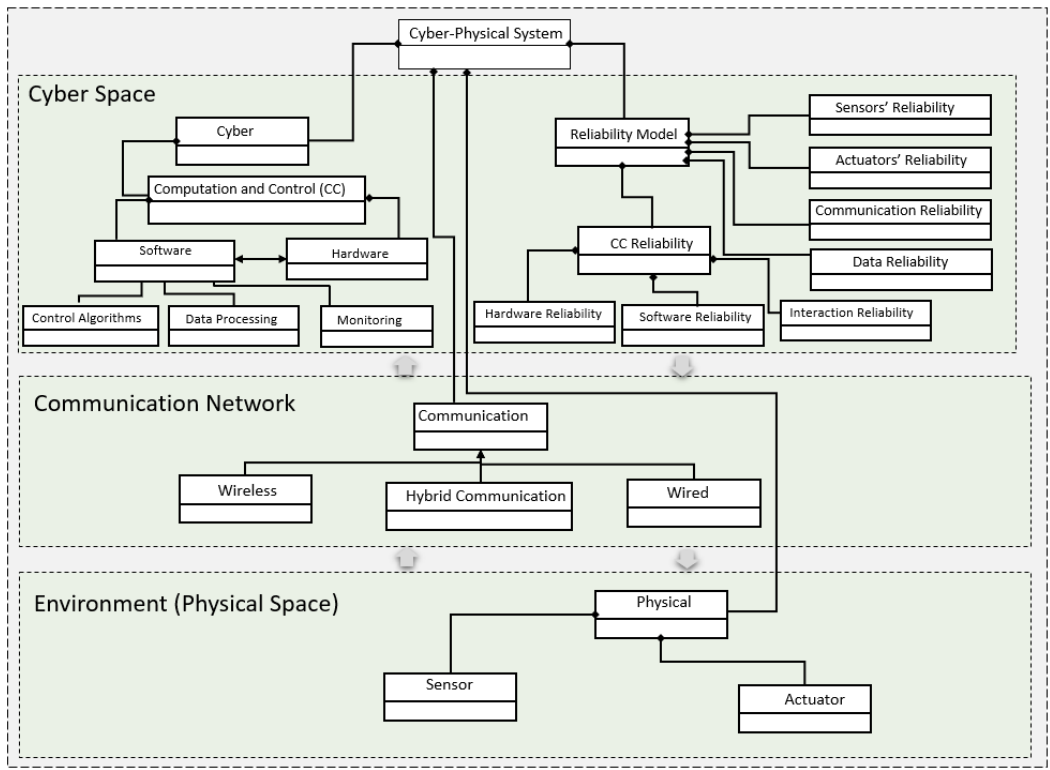

Fig. 1 A generic architecture for reliability modeling and analysis for CPS

$$R_{sys} = P(X_1 \cap X_2 \cap X_3 \cap \ldots \cap X_n)$$
$$= P(X_1)P(X_2|X_1)\,P(X_3|X_1X_2)\ldots P(X_n|X_1X_2\ldots X_{n-1}) \quad (1)$$

Where $R_{sys}$ is system reliability and $P(X_i)$ is the probability that component $i$ works correctly.

When the failure of a system component affects the failure rates of rest of the components (i.e., the life distribution characteristics of the other components change when one component fails), the conditional probabilities in the equation (1) must be taken into account. However, in the case of independent components, equation (1) can be re-written as follows;

$$R_{sys} = P(X_1)P(X_2)P(X_3)\ldots P(X_n) \quad (2)$$

Given that the failure rate for each system component is independent and constant of their usage of time, equation (2) can be written as equation (3).

$$R_{sys} = \prod_{i=1}^{n} R_i \quad (3)$$

Where $R_i$ is the reliability of each independent component $i$ and $n$ is the number of components per operation in the system.

The redundancy is the common technique to improve the reliability of a system [7]. Adding redundant components in the system may lead to high reliability. The redundant component is used when the same working component was failed. In this case, the reliability of the system with redundant components is given in equation (4).

$$R_{sys} = 1 - \prod_{i=1}^{n}(1 - R_i) \quad (4)$$

Where $n$ is the number of redundant components doing the same job.

The reliability of a single component is determined in terms of failure rate as a function of time as mentioned in equation (5).

$$R = e^{-\lambda t} \quad (5)$$

Where $R$ is the reliability of the component and $\lambda$ is the failure rate of the component at time $t$. When we consider the failure rate $\lambda$ as constant mean value (i.e., exponentially distributed time t between failures), the reliabilities as mentioned in equation (3) and (4) are determined on the basis of exponential distribution as shown in equation (5). The mean time between failure (MTBF) of the component can be expressed as equation (6).

$$MTBF = \frac{1}{\lambda} \quad (6)$$

## 2.2 Reliability Modeling for Individual Modules

In this section, we focus on the reliability modeling for individual modules of target CPS. Our CPS architecture has sensor modules, actuator modules, communication modules, and computation and control module.

*1) Sensor Modules*

Let's suppose that a system requires additional sensors for a single task. Let *n* be the total number of sensors, *mn* be the minimum number of sensors required to perform the task. The possible ways to select *mn* from *n* is expressed by n$C$mn. Subsequently, the reliability for any combination can be formulated according to the selection.

Assume that a CPS has a temperature sensor, a humidity sensor, a fire sensor, and a rain sensor to facilitate the inhabitants of smart home [8]. Sensitivity, sampling, noise, frequency, accuracy, and failure rate are considered to parameters to determine sensor performance. However, the failure rate determines the reliability of a sensor as mentioned in equation (7).

$$R_{sensor} = e^{-\lambda_s t} \quad (7)$$

*2) Actuator Module*

The widespread metrics to measure the performance of actuators are speed, durability, torque, force, and operation conditions. The reliability of actuators can be determined by considering the above parameters. Hence, the reliability for actuators can be expressed as equation (8).

$$R_{actuators} = e^{-\lambda_a t} \quad (8)$$

Where, $\lambda_a$ is corresponding failure rate of an actuator.

*3) Communication Network Reliability*

The communication network is the backbone of CPSs which acts as an information carrier. It carries information from physical layer to cyber layer and back to physical layer. There are a number of communication techniques to carry information such as wireless communication, wired communication, and hybrid communication (combination of wired and wireless communication). However, M2M type of communication is widely used in CPSs. M2M architecture has mainly three domain such as M2M device domain, network domain, and application domain [12]. Data from physical layer goes through various levels [2] and converted into information which is used by business processing engines for decision making. M2M gateway is considered communication network node which is used for interfacing among group of networks. This gateway uses various protocols, translators, impedance matching devices, signal isolators and fault isolators to ensure reliability.

*4) Computation and Control Module*

As shown in Fig. 1, the computation and control module is consisting of software and hardware. To measure the reliability of software and hardware, Weibull distribution and Markov process are usually used [9]. The software part of computation and control is independent of time while the hardware module is depending on time. In CPS, the software and hardware are highly intertwined. Therefore, the failure can be divided into three categories such as 1) hardware failure, 2) software failure and 3) software-hardware interaction failure.

The hardware failures can be better handled with Weibull distribution because it is robust to determining distribution depending on parameters such as scale, shape, and location. More specifically, the Weibull is more convenient and reliable for real-time applications. Therefore, the reliability for hardware component is expressed as in equation (9).

$$R_{Hardware}(t) = e^{-\lambda t^\beta} \quad (9)$$

Where, $\lambda$ is combined failure rate for each hardware component, $\beta$ is the shape parameter of the Weibull distribution model. The hardware component in computation and control unit refers to storage and computation devices e.g. CPU etc.

The software reliability growth model (SRGM) is an important model to measure software reliability. The most general SRGM model is the Non-Homogeneous Poisson Process (NHPP) SRGM with mean value function [10]. The base model for SRGM is derived from G-O model [11] and modifies only hypothesis of G-O model. The SRGM based NHPP model is highly recommended for controlling, predicting, and analyzing the reliability of a software. Therefore, the accumulative number of failures at the time $t$ is expressed as:

$$P\{N(t) = n\} = \frac{\{m(t)\}^n}{n!} \exp\{-m(t)\}, n = 0, 1, .., t \geq 0.$$

Where $m(t)$ is the mean value function in a time $t$. The failure intensity function tells the failure occurrence rate at time $t$. The failure intensity function expressed below shows relationship between failure intensity function and mean value function.

$$\lambda(t) = \frac{d}{dt} m(t)$$

In SRGM based NHPP, it is a common assumption that the failure occurrence at a time point $t$ is directly proportional to the anticipated number of failures as expressed below:

$$\lambda(t) = b(t) * \{a(t) - m(t)\}$$

Where $b(t)$ is the fault detection rate function which is time-dependent, and $a(t)$ is the time-dependent fault content function. According to the SRGM based NHPP model, the mean value function $m(t)$ is given as under:

$$m(t) = a * \{1 - e^{-bt}\}, a > 0, b > 0$$

Hence, the failure rate for the software system can be expressed as follows:

$$\lambda(t) = \frac{d}{dt} m(t) = ab \exp[-bt]$$

Therefore, reliability for software component can be written as equation (10).

$$R_{software}(t) = e^{-(m(t+x)-m(t))} = e^{-\lambda_S t} \quad (10)$$

Software interacts with hardware to perform a task [13]. Therefore, software-hardware interaction sometimes leads to software failure. The software-hardware failures occur due to the manipulations in the hardware configurations. It is also possible that the manipulations in software configurations may lead to hardware failure while interacting with hardware. This scenario is called software-hardware interaction failure. We also considered this software-hardware interaction in reliability modeling as shown in Fig.1. Weibull distribution model [9] is proposed to compute the reliability of software-hardware interaction. Hence,

$$\lambda_{SH}(t) = \alpha_{SH*} \beta_{SH*} t^{\beta_{SH}-1} \quad \alpha, \beta > 0, t \geq 0$$

Where $\lambda_{SH}$ is the rate of software-hardware (SH) interaction failure, $\alpha_{SH}$ and $\beta_{SH}$ are scale and shape parameters. We see that the failure rate is proportional to power $\beta_{SH} - 1$ of time $t$.

The mean value function of software-hardware interaction failure will be scale parameter times power shape parameter of time $t$ as shown below:

$$m(t) = \alpha_{SH*} t^{\beta_{SH}}$$

Therefore, the reliability for interaction failure can be expressed as:

$$R_{SH}(t) = e^{-(m(t+x)-m(t))}$$
$$= e^{-(\alpha_{SH*}(t+x)^{\beta_{SH}} - \alpha_{SH*} t^{\beta_{SH}})} \quad (11)$$

Hence, reliability for computation and control module would be formulated in equation (12).

$$R_{CC}(t) = R_{software}(t) + R_{Hardware}(t) + R_{SH}(t) \quad (12)$$

Therefore, reliability estimation for the entire CPS without data reliability can be modeled as equation (13).

$$R_{CPS} = R_{CC} * R_{Actuator} * R_{sensors} * R_{CN} \quad (13)$$

*5) Data Reliability*

Data reliability means the quality characteristics of the data that must be processed by the software system. Data processed in CPS applications can be categorized with (1) internal data processed by the application itself, and (2) data transmitted from other external systems. Their representative examples include data input from various types of sensors, and messages from interactive external systems. All of these data must be taken into account in common way whether data quality or data reliability is appropriate for processing in CPS system. Namely, data reliability is an evaluation of whether the data meets its intended use and can be expressed in terms of accuracy, consistency, and completeness [31].

Especially, data reliability in CPS applications also includes timeliness in addition to these basic quality characteristics. This is a factor that must be available when data is required. These characteristics are shown in Table 1 and can be expressed as shown in equation (14).

$$R_{Data} = D_c + D_a + D_i + D_t \quad (14)$$

Table 1 Critical factors of data reliability on CPS applications.

| Data Quality | Description | Measurement Approach |
|---|---|---|
| Completeness (Dc) | Is there any data gap? | Count the omitted data items based on input specification |
| Accuracy (Da) | Are all the data within the acceptable range? Do data fit into the usual profile? | Count the data items which over-run the range and precision of the specified values |
| Consistency (Di) | Are all the data in the same format? | All data item to be processed should be matched its types and formats with the specified data types and formats |
| Timeliness (Dt) | Do data arrive on time? | Count the data items which violate the specified time limits. |

Therefore, the reliability estimation for the entire CPS with data reliability can be modeled as the equation (15).

$$R_{CPS} = R_{CC} * R_{Actuator} * R_{sensors} * R_{CN} * R_{Data} \quad (15)$$

## 3. RELIABILITY COMPUTATION WITH AN ILLUSTRATIVE EXAMPLE

We assume that a CPS is designed with off the shelf components available in the market. Each CPS component with different implementations and configurations can be selected while using off the shelf components. The components with the highest reliability can be chosen after reliability analysis. We used a dataset from [25] and calculated the reliabilities for each component as shown in Fig.2. The proposed CPS has five sensors (S1, S2, S3, S4, S5) and three actuators (A1, A2, A3) with additional redundant five sensors and three actuators to improve reliability. The proposed system also has two communication networks and two computation and control units instead of one to ensure the reliability. Each module (sensor module, actuator module, communication, and network module, computation and control module) has replicated components. Therefore, equation (4) was used to calculate reliability for parallel redundant components. The reliability for entire CPS, without considering data reliability, was then calculated using equation (13) by selecting highest reliable component from each module. Hence, entire CPS reliability was 0.9538 without considering data reliability.

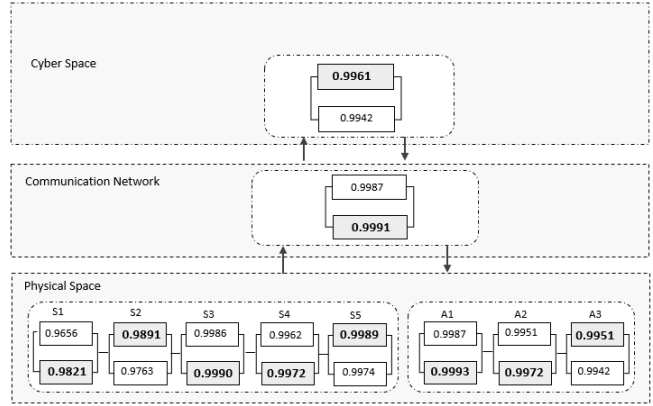

Fig. 2 Component selection for highest reliability

## 4. RELATED WORK

This section provides related work about the reliability modeling of CPS, and then presents a survey for interaction (software-hardware and hardware-software interaction) reliability. Table 2 shows the summary of related work with their focus and approach overview.

### 4.1 Reliabiliyy Modeling for CPS

For safety-critical systems, continuous availability is a firm requirement and software reliability is one of important components of continuous system availability. A number of researchers considered reliability for CPSs. Leon Wu *et al*. [6] presented a failure analysis and reliability estimation framework to benchmark reliability for CPSs. The proposed framework provides methods and metrics for reliability analysis, reliability estimation and operational availability to benchmark CPS reliability. This framework provides a general but accurate overview for representing CPS reliability. The authors introduced three additional reliability metrics to cover specific evaluation scenarios.

Mavrogiorgou *et al*. [26] proposed a generic approach for measuring reliability in medical CPSs. The proposed framework has four stages: 1) the CPS modeling, 2) evaluation environment, 3) failure analysis, and 4) reliability estimation. In CPS modeling stage, the CPS is proposed to be modeled according to the application domain through a domain-specific language. In the evaluation environment stage, after CPS modeling, the evaluation environment is selected. In evaluation environment, failure data is collected, and reliability tests are implemented. After selection of environment, failure analysis begins, which includes failure detection and analysis, along with domain knowledge and heuristics. The reliability of whole system can be improved when domain knowledge data is well managed [27]. The content relationship among multiple hazard analysis techniques can also be used to detect failure [28]. Lastly, after failure detection and analysis, reliability of the system is computed using different reliability estimation metrics.

Nannapaneni *et al.* [29] proposed a framework in order to formulate CPS reliability evaluation as a dependency issue derived from functional requirements, software components, and physical system dependencies. The proposed framework is argued to analyze CPS reliability in the early design phase. Koc et al. [32] proposed a study that models the reliability for CPSs. The authors model reliability for software, hardware, sensors, and actuators. However, the proposed architecture is a simple one that does not cover all kinds of reliabilities such as data reliability.

Some researchers have proposed reliability benchmark [4] for specific CPSs, i.e. wireless sensor networks and wind plant. Wu et al. [5] have presented a hierarchical model for reliability analysis with the focus on the integrated modular avionics system. However, it is domain specific framework which cannot be applied for developing general CPSs considering application domains.

## 4.2 Interaction Reliability

The equation (13) is one of the general models to compute reliability for entire cyber-physical systems. However, different researchers have categorized the software-hardware interaction failures in two ways [13]: 1) Hardware-driven software interaction failures, which are software failures caused by hardware, 2) software-driven hardware failure is the failure caused by faulty software. In this section we survey the literature that considered the above two cases for computing software-hardware interaction reliability.

### *4.2.1 Software Failures caused by Hardware*

It is the type of failure caused due to a fault in hardware or changes in hardware configuration. Teng *et al.* [14] have argued that software systems fail due to hardware component fails, or software component fails or hardware-software interaction failure. The authors have used Weibull distribution model for hardware component and NHPP for software components to estimate reliability. The failure distribution of hardware-software interaction was modeled using Markov chain. The authors computed system reliability by taking product of independent software failure, independent hardware failure, and hardware-software interaction failure probability. However, the Markov Improving System Reliability for Cyber-Physical Systems chain model has a number of limitations such as some critical information e.g. the number of cumulative detected failures, test time, and software release time cannot be used during system testing.

Sinha Roy *et al.* [15] presented a reliability framework to estimate reliability for Phasor Measurement Units. The authors used Weibull distribution model for hardware component failure, NHPP to model the software components failure, and Markov chain to model the hardware-software interaction failures. However, the authors used Monte Carlo simulation to validate their proposed model.

Kanoun *et al.* [16] have proposed an approach which employed Generalized Stochastic Petri Net (GSPN) and Markov chain for modeling dependability of a fault-tolerant system by considering hardware and software dependency. The authors have considered two types of interactions: 1) interaction among internal system components and 2) interaction with external distributed system components. The authors have used Petri net for dependability analysis of software, hardware, and their interaction. It was argued that it is difficult to analyze a graph with a thousand system sates generated by GSPN.

Sumita *et al.* [17] proposed a novel availability/reliability analysis approach which considered system failure as a multivariate stochastic process using Markov chain and matrix Laguerre transformation. This approach assumes that hardware failures are independent of software failures. The authors assumed that hardware subsystem was composed of an independent alternating renewal process that has exponential up-times and general downtimes. They argued that there would be two possibilities if hardware-related failures lead to software failures: 1) software repair would complete without interruption from hardware failure, 2) hardware failure would interrupt software repair operation. The authors modeled both cases using stochastic process and reliability of system was computed using matrix laguerre transformation.

### *4.2.2 Hardware Failures caused by Software*

Those are hardware failures caused due to a fault in the software. The researchers [18-24] have highly used system modeling approach to address software-hardware interaction. Majority of these approaches used functional failure identification and propagation (FFIP) or its variations to reflect hardware failures caused by faulty software.

Xiaoxu *et al.* [18] investigated the propagation and effects of faults in safety-critical components in nuclear hybrid energy systems. The investigation results were used to design an online monitoring system. The online monitoring system had enough capability to detect and analyze faults during system operations. The authors used integrated system failure analysis, an effective method that analyzes hardware-software faults during design phase to detect failures. First, system components required for integrated system failure analysis were built. Then, fault propagation was implemented based on the acceptance criteria derived for the design of online monitoring system. The results were then analyzed to detect the faults during system operation.

David *et al.* [19] addressed hardware-software interaction failures by proposing FFIP framework. The failure propagation in the combined software-hardware system was addressed through high-level system modeling. The authors are believed to be the first to introduce FFIP framework for system level analysis. The FFIP approach combines both system modeling and behavior simulation techniques in order to analysis the failures in early design phase. The approach was applied to redundancy management system of a NASA space shuttle to evaluate nonlinear failure propagation within system.

Chetan et al. [20] argued that the proposed FFIP [19] framework is only efficient to detect electromechanical faults. It can not work efficiently in detecting faults residing in cross-domain functionalities. In order to cope with this problem, the authors introduced the Integrated System Failure Analysis (ISFA) approach to identify and analyze faults in cross-domain functionalities. They proposed Failure Propagation and Simulation Approach (FPSA) to model fault propagation and its effects analysis method using UML. The FPSA is actually an extension of FFIP.

Table 2 Summary of related work with their domain and overview

| Proposed Reliability Models | Domain | Technique Overview |
|---|---|---|
| Leon Wu et al. [6] | General CPS | Presents a framework for benchmarking reliability of general cyber-physical systems |
| Mavrogiorgou et al. [26] | Medical | Presents a generic Approach for Capturing Reliability in Medical Cyber-Physical Systems |
| Nannapaneni et al. [29] | General | This study addresses a dependence problem derived from the software component dependences, functional requirements and physical system dependences |
| Koc et al. [32] | General | The authors proposed a general reliability model to measure the reliability of whole system. However, the proposed architecture is a simple one that does not cover all kinds of reliabilities such as data reliability. |
| Wu et al. [5] | Avionics | This is a hierarchical model for reliability analysis with the focus on the integrated modular avionics system. |
| Teng et al. [14] | General CPS | The authors presented a reliability model to measure reliability for software-hardware interaction reliability. |
| Sinha Roy et al. [15] | Electrical | A reliability framework to estimate reliability for Phasor Measurement Units |
| Kanoun et al. [16] | Cross-domain | A Generalized Stochastic Petri Net (GSPN) and Markov chain framework for modeling dependability of a fault-tolerant system by considering hardware and software dependency |
| Sumita et al. [17] | Embedded system | A hardware-software reliability model where both the hardware and software subsystems are subject to random failures. |
| Xiaoxu *et al.* [18] | Nuclear energy system | This study presents integrated system failure analysis, an effective method that analyzes hardware-software faults during design phase to detect failures. |
| David et al. [19] | NASA space shuttle | This study presents hardware-software interaction failures by proposing FFIP framework. It cannot work efficiently in detecting faults residing in cross-domain functionalities. |
| Chetan et al. [20] | Cross-domain | This study presents an Integrated System Failure Analysis (ISFA) approach to identify and analyze faults in cross-domain functionalities. Failure Propagation and Simulation Approach (FPSA) is proposed in this study to model fault propagation and its effects analysis method using UML. |
| Nikolaos *et al.* [21] | Nuclear reactor | Presents an alternative flow path to evade failure propagations through functional failure identification and propagation (FFIP). |
| Nikolaos et al. [22] | Nuclear Power Plant | This study presents hierarchical function fault detection and identification (HFFDI) approach to address the fault propagation. The presented framework combined machine learning with FFIP to achieve better performance |
| Seppo et al. [23] | Nuclear reactor | This study is a risk analysis approach to identify fault propagation paths residing in cross-domain functionalities. This approach is an extension of FFIP. |
| Tumer et al. [24] | NASA space shuttle | This study analyzes failures in the combined software-hardware system. It is a system-level modeling and reasoning approach to identify failure propagation paths. |
| OURS | General Benchmarking reliability model | We present a general framework for capturing reliability in CPS. We consider |

Nikolaos *et al.* [21] found some drawbacks in FFIP framework based system models. The authors were of the opinion that using different techniques for modeling the same system can yield different results. To cope with this drawback, an alternative flow path was proposed to evade failure propagations. This approach is inspired by [24] and analyze risk of failure propagation through FFIP. The authors applied their approach to a boiling water nuclear reactor in a Simulink environment to shows its effectiveness.

The authors [21] further improved their previous approach and proposed a hierarchical function fault detection and identification (HFFDI) approach to address the fault propagation [22]. The proposed framework combined machine learning with FFIP for better performance. Machine learning techniques were used to detect and identify faults from the historical data while the functional decomposition of the system was done through FFIP. This framework was applied to a nuclear power plant system and compared failure analysis of HFFDI with Fault Detection and Identification (FDI) for same case study. The authors observed that HFFDI gave better performance than its counterpart.

Seppo *et al.* [23] presented a risk analysis approach to identify fault propagation paths residing in cross-domain functionalities. This approach is an extension of FFIP [19]. The authors used FFIP instead of traditional system decomposition and risk assessment techniques such as FTA, and FMEA etc. At the early design phase, the system was expressed with syntax and semantic that was able to describe the fault propagation throughout system and particularly cross-domain functionalities. The proposed approach was applied to a boiling water nuclear reactor to shows its applicability. The results revealed the capability of proposed approach to handle several fault propagations paths in a single scenario for hazard identification at early design phase.

Tumer *et al.* [24] extended FFIP [19] in order to analyze failures in the combined software-hardware system. The proposed approach is a system-level modeling and reasoning approach to identify failure propagation paths. The proposed approach works in five steps: 1) developing a system layout by identifying functions, components, and flow taxonomies and identify various configuration models to reflect system design, 2) identifying different failure modes for components 3) identifying components and subcomponents that can act as guard conditions, 4) find scenarios of one or more initiating failures, 5) evaluating different scenarios against predefined set of rules for alternate flow path.

## 5. CONCLUSION

Cyber-Physical systems have grown significantly in various application domains. The CPSs systems are widely believed as safety critical systems. For safety-critical systems, continuous availability is a necessity and software reliability is one of important components of continuous system availability. In this work in progress, we present an architecture to capture reliability for CPSs. The proposed architecture has the reliability characteristics for sensor, actuator, communication and network, data, software and hardware, and interaction (software-hardware, hardware-software). Using the reliability models presented, we are working to measure the reliability for Automatic Incident Detection System (AIDS), as one of CPS.

In our future work, we are planning to determine reliability for entire CPS using equation (15) as defined in our proposed framework. Also, we are considering to developing safety model in quantitative manner.

## 6. ACKNOWLEDGMENT

This research was supported by the NRF of Korea funded by the MSIT (NRF-2017M3C4A7066479).